\documentclass{article}
\usepackage{spconf}
\usepackage{amsmath}
\usepackage{graphicx}
\usepackage{array}
\usepackage{url}
\usepackage{multirow}
\usepackage{balance}
\usepackage{bbm}
\usepackage{cite}
\usepackage{balance}
\usepackage{xcolor}

\title{Self-supervised Speaker Recognition with Loss-gated Learning}

\name{Ruijie Tao$^{1}$, Kong Aik Lee$^{2}$, Rohan Kumar Das$^{3}$, Ville Hautamäki$^{1,4}$ and Haizhou Li$^{1,5}$ 
\thanks{This work is supported by 1) Human-Robot Interaction Phase 1 (Grant No. 1922500054), National Research Foundation (NRF) Singapore under the National Robotics Programme 2) Neuromorphic Computing (Grant No. A1687b0033), the Singapore Government's Research, Innovation and Enterprise 2020 plan (Advanced Manufacturing and Engineering domain) 3) Human Robot Collaborative AI for AME (Grant No. A18A2b0046), NRF Singapore}}

\address{
  $^{1}$Department of Electrical and Computer Engineering, National University of Singapore, Singapore\\
  $^{2}$Institute for Infocomm Research, A$^\star$STAR, Singapore~~~~~~$^{3}$Fortemedia Singapore, Singapore\\
  $^{4}$University of Eastern Finland, Finland~~~~~~$^{5}$The Chinese University of Hong Kong, Shenzhen, China}
 
\begin{document}
\topmargin=0mm
\maketitle

\begin{abstract}
In self-supervised learning for speaker recognition, pseudo labels are useful as the supervision signals. It is a known fact that a speaker recognition model doesn't always benefit from pseudo labels due to their unreliability. In this work, we observe that a speaker recognition network tends to model the data with reliable labels faster than those with unreliable labels. This motivates us to study a loss-gated learning (LGL) strategy, which extracts the reliable labels through the fitting ability of the neural network during training. With the proposed LGL, our speaker recognition model obtains a $46.3\%$ performance gain over the system without it. Further, the proposed self-supervised speaker recognition with LGL trained on the VoxCeleb2 dataset without any labels achieves an equal error rate of $1.66\%$ on the VoxCeleb1 original test set. Code has been made available at: \textcolor{magenta}{\url{https://github.com/TaoRuijie/Loss-Gated-Learning}}.

\end{abstract}
 \vspace{-3mm}
\section{Introduction}
 \vspace{-2mm}
\label{sec:1}
\begin{figure*}[!ht]
    \centering
    \includegraphics[width=\linewidth]{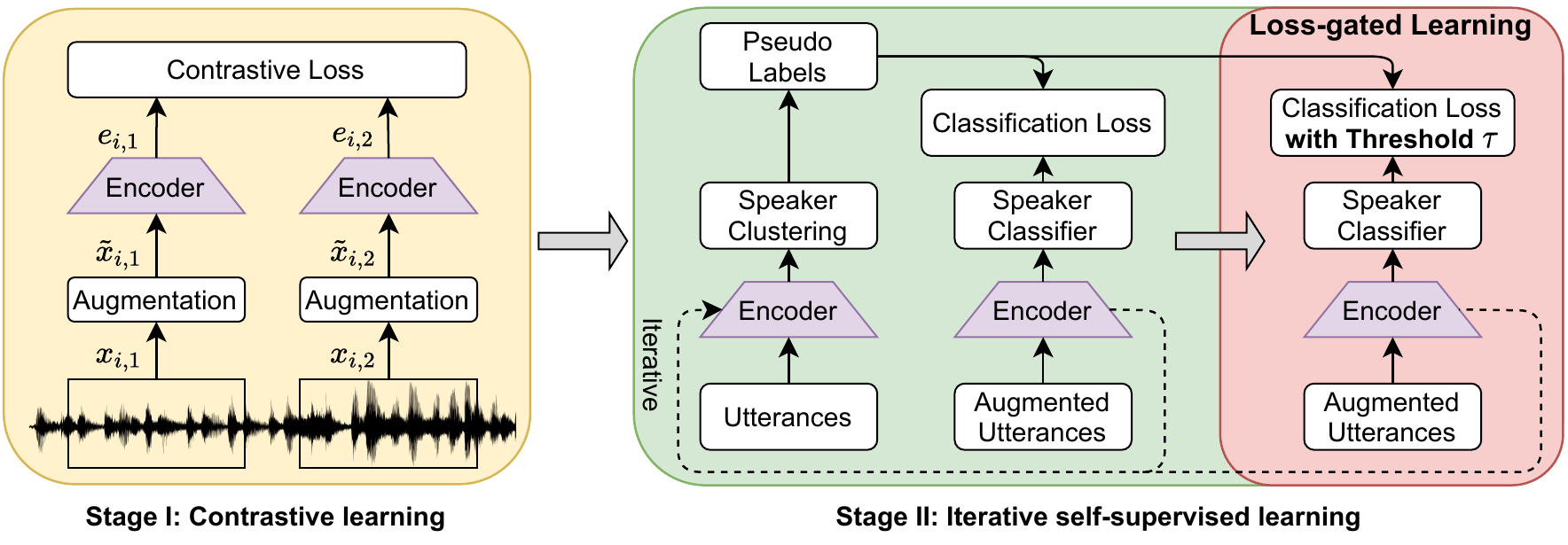}
    \caption{Framework of self-supervised speaker recognition with loss-gated learning.}
    \label{fig:LGL}
\end{figure*}
Speaker recognition aims to recognize persons from their voices~\cite{Tomi, lee2011joint, APSIPA_realism, zhou2021language}. Over the last decade, speaker recognition models trained via supervised learning have achieved remarkable performance~\cite{snyder2018x, chung2020defence, desplanques2020ecapa, Two_decades, liu2022neural,Tianchi_icassp22}. However, these methods usually require a large set of data with manually annotated speaker labels. The creation of such annotated data is not only immensely costly, but also laborious. Self-supervised learning doesn't require such speaker labels~\cite{stafylakis_is19,nagrani2020disentangled, chung2020seeing}, that opens up the opportunity to leverage the abundant unlabeled speech resources. 

The state-of-the-art self-supervised speaker recognition system consists of two stages~\cite{cai2021iterative, thienpondt2020idlab}. In Stage I, we solve the contrastive pretraining task through SimCLR~\cite{chen2020simple, cai2021iterative} or MoCo~\cite{ he2020momentum,xia2021self}, then the speaker encoder can learn the meaningful speech representation. Meanwhile, various loss functions are proposed to set contrastive targets~\cite{chung2020seeing,mun2020unsupervised,zhang2021contrastive,huh2020augmentation}. However, the performance at the Stage I is limited due to the lack of speaker identity information. Recently, an iterative Stage II~\cite{cai2021iterative, thienpondt2020idlab} is proposed to address this issue, where a clustering algorithm is applied to generate the pseudo labels for each utterance based on the learnt representation. With the pseudo labels, the network is trained with a classifier in a supervised manner. This process is repeated several times to improve the speaker encoder.

The state-of-the-art studies take the pseudo labels for fully supervised classification. Therefore, the quality of classification in the Stage II decides the upper bound of self-supervised speaker recognition~\cite{cai2021iterative, thienpondt2020idlab}. Usually, the pseudo labels include massive unreliable labels~\cite{caron2018deep}. Such unreliable pseudo labels will adversely affect the performance of the encoder, that highlights the importance of having an effective and reliable selection of pseudo labels~\cite{rizve2020defense}, which is similar with label smoothing~\cite{li2020regularization}. 

In this work, we hypothesize that neural networks model the data with reliable labels faster than the those with unreliable labels. Specifically, considering one utterance at the time, if the forward pass yields a low loss value, we can refer its pseudo label to be reliable, whereas in the case of high loss value, its pseudo label is unreliable. We design a toy experiment to validate our hypothesis. To this end, we propose a \emph{loss-gated learning} (LGL) method to effectively select the data with reliable labels. A threshold is involved to retrain the data with small loss. Only the filtered data are then used to update the network. We believe the proposed LGL is capable of selecting the reliable labels to contribute towards an improved performance. In summary, we make the following contributions:
\begin{itemize}
\vspace{-2mm}
\item We confirm that neural networks fit the reliable pseudo labels faster than the unreliable ones in self-supervised speaker recognition.
\vspace{-3mm}
\item Based on our finding, a LGL strategy is proposed to effectively select the reliable pseudo speaker labels. We compare LGL to the baseline method by experiments.
\vspace{-3mm}
\end{itemize}
\section{Baseline: Two-stage architecture}
\vspace{-2mm}
\label{sec:2}
We now describe our baseline two-stage architecture. As shown in Fig.~\ref{fig:LGL}, a contrastive learning task is employed to train a speaker encoder. Then we obtain the pseudo labels by clustering and train a classification network iteratively.

\subsection{Stage I: Contrastive Learning}
In Stage I, we design a self-supervised pretraining task from the \emph{simple contrastive learning} (SCL)~\cite{chen2020simple,cai2021iterative}. For each mini-batch, we randomly select $N$ unlabelled utterances $x_1, \cdots, x_N$. As shown in Fig.~\ref{fig:LGL}, we randomly consider two non-overlapping segments $x_{i,1}$ and $x_{i,2}$ with the same length for each utterance $x_i$. Then for each of these segments, we apply the stochastic noise augmentation to get the augmented segments $\widetilde{x}_{i,1}$ and $\widetilde{x}_{i,2}$. These segments are fed into the speaker encoder $ f(\cdot)$ to obtain the speaker embeddings $ e_{i,j} = f(\widetilde{x}_{i,j}) $, where $i \in \{ 1, \cdots, N\}$ and $j \in \{ 1, 2\}$.
We assume that each utterance contains only one speaker. The segments drawn from the same utterance share the same speaker identity, and therefore they form the positive pairs. On the other hand, we form negative pairs from the segments drawn from the different utterances. It is noted that by considering the batch size and the large number of speakers~\cite{zhang2021contrastive}, the probability of selecting a false-negative pair is very low. In order to attract the positive pairs and repel the negative pairs, we define the contrastive loss for each positive pair against all the negative pairs as~\cite{chen2020simple}:
\begin{equation}
    \label{e1}
    l_{i,j} = -\log \frac{\exp(\cos(e_{i,1}, e_{i,2}))}{\sum_{k=1}^{N}\sum_{l=1}^{2}\mathbbm{1}_{\substack{{k \neq i} \\ {j \neq l}}}\exp(\cos(e_{i,j}, e_{k,l}))}
\end{equation}
The loss function for each mini-batch is then given by:
\begin{equation}
    \label{e2}
    L_{\mathrm{scl}} = \frac{1}{2N} \sum_{i=1}^{N} \sum_{j=1}^{2} l_{i,j}
\end{equation}
Notice that the function $\cos(\cdot, \cdot)$ denotes the cosine similarity and we do not set the temperature parameter. By minimizing this loss function, the speaker encoder learns the utterance representations that discriminate positive pairs against negative pairs.

\subsection{Stage II: Iterative Self-Supervised Learning}
Stage II can be viewed from Fig.~\ref{fig:LGL}. First, we use the speaker encoder trained in Stage I as the initial model to extract the speaker embeddings for each utterance. Based on these embeddings, the $k$-means clustering~\cite{lloyd1982least} is performed to produce pseudo speaker labels. We interpret that the utterances in the same cluster share the same identity information. Next, we retrain the speaker encoder using these pseudo labels. The classification layer contains one fully connected layer. The \emph{additive angular margin softmax} (AAM-softmax) loss~\cite{deng2019arcface} is used as the loss function. We repeat both these steps for several iterations until the system converges. It is noted that the encoder trained for speaker classification in each iteration will be used to generate embedding vectors for clustering in the next iteration.

\section{Loss-gated Learning}
\label{sec:3}
First, we make the definition of the reliable and unreliable pseudo labels in Stage II. We can obtain an optimal one-to-one mapping between ground-truth speaker labels and clustered pseudo labels through the Hungarian algorithm~\cite{kuhn1955hungarian}. If the ground-truth label of the utterance is the same as the mapped pseudo label, we define this data has the reliable pseudo label. Otherwise, it has the unreliable pseudo label. It is noted that we  cannot provide these reliability information to our self-supervised framework.

Traditionally, the baseline system performs iteratively training with all pseudo labels in Stage II. So the erroneous information from those unreliable pseudo labels will also propagate iteratively, which drops the performance~\cite{rizve2020defense, cai2021iterative}. This prompts us to use the data with reliable pseudo labels only for training to learn the accurate speaker identity.

The question is how to select the reliable pseudo labels effectively. The study in~\cite{wang2019symmetric} shows that some classes were easier to learn and converge faster than other classes for image classification task. Similarly, the learning ability and convergence speed of data with reliable pseudo speaker labels should be different from those with unreliable labels. To validate this hypothesis, We evenly select 1,000 utterances from 10 speakers and perform $k$-means clustering based on the speaker encoder ($k$=10) as a toy experiment. The network is then trained to distinguish these speaker labels.

\begin{figure}[!htt]
    \centering
    \vspace{-2mm}
    \includegraphics[width=\linewidth]{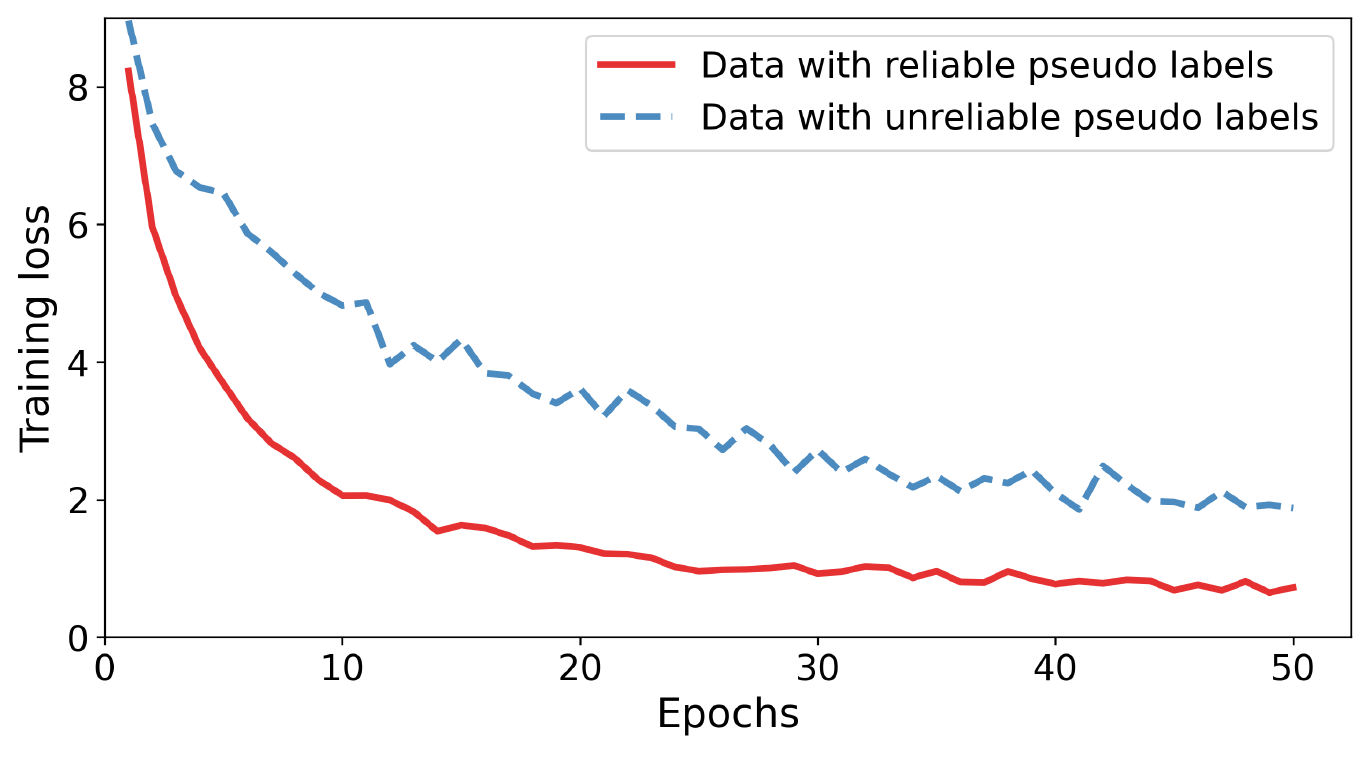}
    \vspace{-5mm}
    \caption{The training loss on 1,000 utterances trained by the pseudo labels.}
    \label{fig:Toy}%
    \vspace{-3mm}
\end{figure}

The average training loss curve is shown in Fig.~\ref{fig:Toy}. Although no reliability or unreliability information is provided to the network during training, we observe the training loss is pushed faster to lower levels with reliable labels (the red curve) vs unreliable labels (the blue curve) automatically. This indicates that fitting reliable labels is easier than fitting unreliable labels. In other words, the model tends to learn useful information from the reliable labels first and then extracts misleading information from unreliable labels. 

Based on that, we proposed the \emph{loss-gated learning} (LGL) strategy. As shown in Fig.~\ref{fig:LGL}, after the pseudo label supervised learning in Stage II, we propose to select reliable data via continuing to train the network with LGL, instead of entering the next iteration directly. The LGL trains the encoder and classifier using the same set of pseudo labels, as in the original supervised learning, with a slight modification to the loss function. Specifically, we introduce a threshold $\tau$ into the loss function as follows:
\begin{equation}
    \label{e4}
    L_{spk} = \sum_{i=1}^{N}l_{i}\mathbbm{1}_{l_i < \tau}
\end{equation}
where $l_i$ is the training loss for a data point. The assumption is that, after training some epochs, data with a small loss is more likely to be more reliable compared to these with larger losses. Thus, LGL only uses the data with a small loss to update the parameters of the network. We train the encoder together with the classification layer until the system performs the best. Then we utilize the trained encoder to do the clustering for the next iteration. To summary, both our proposed and the baseline method have two stages, LGL works on the Stage II to solve the unreliable pseudo label issue.

\section{Experimental setup}
\label{sec:4}
We use the development set of VoxCeleb2~\cite{Voxceleb2} for training. Original, Extended and Hard VoxCeleb1 test sets (Vox\_O, Vox\_E and Vox\_H)~\cite{Voxceleb, Voxceleb2} are used for evaluation. No speaker label information is employed. The emphasized channel attention, propagation and aggregation in time-delay neural network  (ECAPA-TDNN) in~\cite{desplanques2020ecapa} is used as a speaker encoder. The channel size is set at 512. The input is the 80-dimensional log mel-spectrogram from the speech segments. On the other hand, the output is the 192-dimensional speaker embedding.

During the training process, the mini-batch size is set at 256. The network parameters are optimized by Adam optimizer~\cite{2015Adam}. The MUSAN~\cite{MUSAN} and RIRs~\cite{RIRS} datasets are used for data augmentation. The test set provides a set number of pairs for validation and each pair contains two utterances. The cosine similarity score between the speaker embeddings of the given utterance pairs is calculated. The performance metric is the equal error rate (EER).

In the Stage I, the discriminator training in~\cite{huh2020augmentation} is added to build a robust encoder. The initial learning rate is 0.001 that decreases 5\% in every 5 epochs. The duration of the input utterance is 1.8 seconds.

In the Stage II, we employ the $k$-means algorithms without data augmentation by faiss library\footnote{https://github.com/facebookresearch/faiss}~\cite{caron2018deep}. Following the previous work~\cite{cai2021iterative, cai2021dku}, we set the number of clusters as 6,000. Specifically, by calculating the total within-cluster sum of squares for the clustering outputs with a range of speaker numbers, the ‘elbow’ of the curve can be set as the number of the clusters. During training the network with pseudo labels, we set the margin as 0.2 and the scale as 30 in the AAM-softmax loss and fix the learning rate as 0.001. In addition, the duration of the input utterance is 3 seconds. In LGL, we set the hyper-parameter $\tau$ as \{1, 3, 3, 5, 6\} in the five iterations to guarantee that the number of selected data increases by at least 10\% in each iteration. In the last iteration, we extend the channel size of the speaker encoder to 1024 to build a robust system.

\section{Results and Analysis}
\label{sec:5}
\subsection{Proposed System Comparison to Existing Works}
\label{sec:5.1}
We compare the proposed framework with the existing methods in Table~\ref{tab:SOTA}. On Vox\_O set, we implement the two-stage architecture that achieves an EER of $7.36\%$ and $3.09\%$ in Stage I and Stage II, respectively. This proves the robustness of our baseline. Meanwhile, our proposed self-supervised speaker recognition system with LGL obtains an EER of $1.66\%$ on Vox\_O set, which outperforms the best existing method~\cite{thienpondt2020idlab} by $20.95\%$. It also outperforms the previous work~\cite{cai2021iterative} by $45.77\%$ and $42.77\%$ on Vox\_E and Vox\_H sets. 

\begin{table}[t!]
  \caption{Performance of self-supervised speaker recognition with and without LGL in EER (\%). A comparison to other existing works is shown as well.}
  \vspace{2mm}
  \label{tab:SOTA}
  \begin{tabular}{p{0.5cm}<{\centering}p{3.2cm}<{\centering}p{0.9cm}<{\centering}p{0.9cm}<{\centering}p{0.9cm}<{\centering}}
    \hline
    \textbf{Stage} & \textbf{Method} & \textbf{Vox\_O} & \textbf{Vox\_E} & \textbf{Vox\_H}\\
    \hline
    \multirow{8}{*}{I} & Nagrani et al.~\cite{nagrani2020disentangled} & $22.09$ & - & - \\
                       & Chung et al.~\cite{chung2020seeing} & $17.52$ & - & - \\
                       & Inoue et al.~\cite{inoue2020semi} & $15.26$ & - & - \\
                       & Huh et al.~\cite{huh2020augmentation} & $8.65$ & - & - \\
                       & Zhang et al.~\cite{zhang2021contrastive} & $8.28$  & - & - \\
                       & Xia et al.~\cite{xia2021self} & $8.23$ & - & - \\
                       & Mun et al.~\cite{mun2020unsupervised} & $8.01$ & - & - \\
                       & \textbf{Ours} & $7.36$ & $7.90$ & $12.32$\\
    \hline
   \multirow{4}{*}{II} & Cai et al.~\cite{cai2021iterative} & $3.45$ & $4.02$ & $6.57$  \\
                       & \textbf{Ours w/o LGL} & $3.09$ & $3.81$ & $6.32$ \\
                       & Thienpondt et al.~\cite{thienpondt2020idlab} & $2.10$ & - & -  \\
                       & \textbf{Ours with LGL} & $\textbf{1.66}$ & $\textbf{2.18}$ & $\textbf{3.76}$\\
    \hline
  \end{tabular}
\end{table}

\subsection{Impact of LGL: Iterative Analysis}
\label{sec:5.2}
We summarize the performance in each iteration with and without LGL on Vox\_O set in Table~\ref{tab:LGL}. From the table, it can be observed that LGL can quickly promote the system performance in each iteration and finally brings $46.3\%$ improvement compared with the baseline. In addition, LGL is found to be more effective in the beginning since there are more unreliable data.

\begin{table}[h!]
  \vspace{-4mm}
  \caption{Impact of LGL on performance in EER (\%) and comparison to baseline without LGL on Vox\_O set in Stage II.}
  \vspace{2mm}
  \label{tab:LGL}
  \begin{tabular}{p{2.4cm}<{\centering}p{0.7cm}<{\centering}p{0.7cm}<{\centering}p{0.7cm}<{\centering}p{0.7cm}<{\centering}p{0.7cm}<{\centering}}
    \hline
    \textbf{Iteration-\#} & $\textbf{1}$ & $\textbf{2}$ & $\textbf{3}$ & $\textbf{4}$ & $\textbf{5}$ \\
    \hline
    \textbf{W/o LGL} & $4.92$ & $4.00$ & $3.68$ & $3.22$ & $3.09$ \\
    \textbf{With LGL} & $3.52$ & $2.41$ & $2.07$ & $1.95$ & $\textbf{1.66}$ \\
    \hline
  \end{tabular}
  \vspace{-6mm}
\end{table}

\subsection{Ablation Study}
To study the robustness of our LGL and validate our hypothesis, we do the following experiments in Stage II and evaluate the performance on Vox\_O set.

\vspace{-4mm}
\subsubsection{Robustness to Number of Clusters}
Table~\ref{tab:Number_C} reports the performance with different number of clusters in the Iteration-1. It is observed that the reasonable setting for the number of clusters will not effect the performance too much for both the baseline and our proposed LGL method. Meanwhile, our LGL can bring significant and stable improvements for all the given number of clusters.

\begin{table}[ht]
  \vspace{-2mm}
  \caption{Impact of LGL on performance in EER (\%) with different number of clusters in the Iteration-1.}
  \vspace{3mm}
  \label{tab:Number_C}
  \begin{tabular}{p{2cm}<{\centering}p{0.8cm}<{\centering}p{0.8cm}<{\centering}p{0.8cm}<{\centering}p{0.8cm}<{\centering}p{0.8cm}<{\centering}}
    \hline
    \textbf{\# Clusters} & $\textbf{3,000}$ & $\textbf{4,500}$ & $\textbf{6,000}$ & $\textbf{7,500}$ & $\textbf{9,000}$ \\
    \hline
    \textbf{W/o LGL} & $5.29$ & $4.89$ & $4.92$ & $5.05$ & $5.31$ \\
    \textbf{With LGL} & $3.49$ & $3.35$ & $3.52$ & $3.48$ & $3.71$ \\
    \hline
  \end{tabular}
  \vspace{-4mm}
\end{table}

\subsubsection{Robustness to Threshold of LGL}
Then we study the robustness to threshold $\tau$ of LGL in the Iteration-1 as reported in Table~\ref{tab:study}. Compared with the baseline method without threshold ($\tau = +\infty$), all threshold settings from 1 to 5 can bring significant improvement. This proves LGL is relatively robust to this hyperparameter.
\begin{table}[ht]
  \vspace{-4mm}
  \caption{Robustness to threshold in the Iteration-1.}
  \vspace{2mm}
  \label{tab:study}
  \begin{tabular}{p{2.4cm}<{\centering}p{0.5cm}<{\centering}p{0.5cm}<{\centering}p{0.5cm}<{\centering}p{0.5cm}<{\centering}p{0.5cm}<{\centering}p{0.6cm}<{\centering}}
    \hline
    \textbf{Threshold ($\tau$)} & $\textbf{1}$ & $\textbf{2}$ & $\textbf{3}$ & $\textbf{4}$ & $\textbf{5}$ & $+\infty$ \\
    \hline
    \textbf{EER} & \textbf{3.52} & 3.77 & 3.74 & 4.10 & 4.14 & 4.92 \\
    \hline
  \end{tabular}
\end{table}
\vspace{-4mm}
\subsubsection{Post-analysis of Clustering Performance}
\begin{figure}[!htt]
    \vspace{-0.1mm}
    \centering
    \includegraphics[width=1\linewidth]{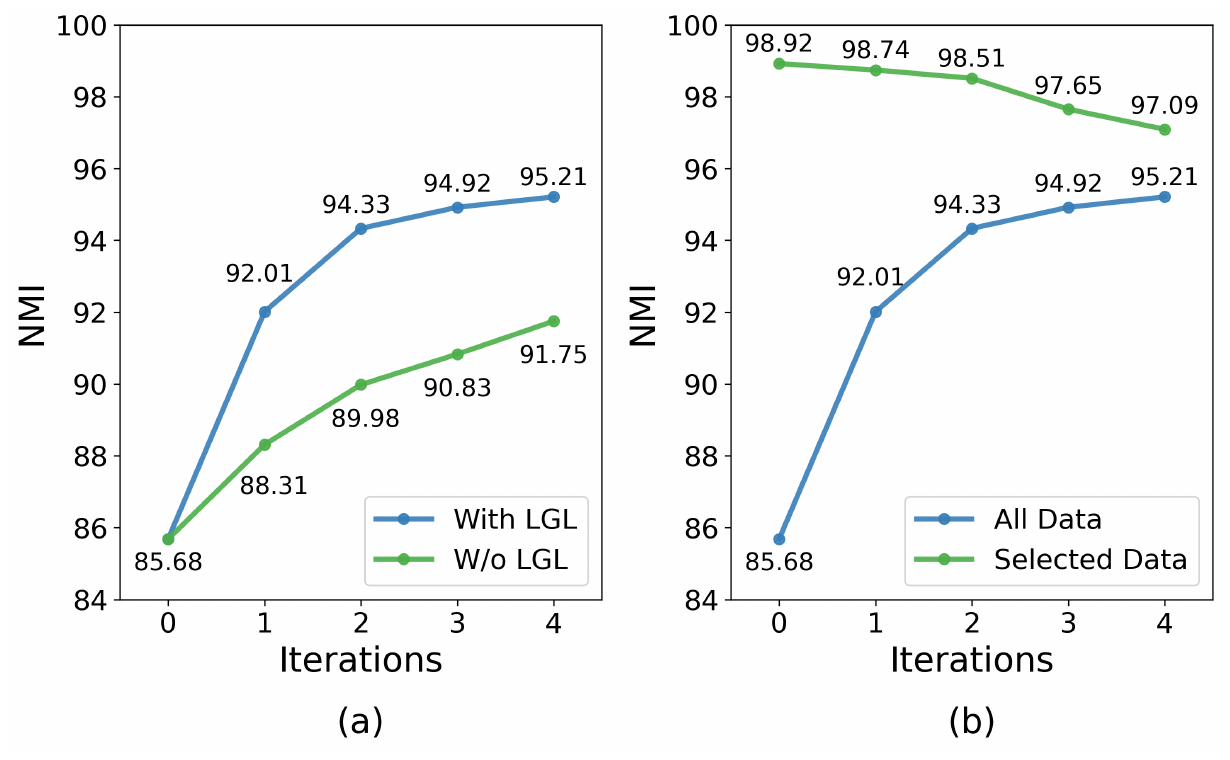}
    \vspace{-4mm}
    \caption{(a) NMI for the system with and without LGL. (b) NMI of the selected data and all data for the system with LGL.}
    \label{fig:NMI}%
    \vspace{-4mm}
\end{figure}

Now we evaluate the clustering performance by normalized mutual information (NMI). The higher NMI indicates a better clustering result. First, we compare the NMI in each iteration with and without LGL in Fig.~\ref{fig:NMI} (a), LGL improves NMI by a large margin and leads to a better clustering result. Then for self-supervised speaker recognition system with LGL, in the Fig.~\ref{fig:NMI} (b), we compare the NMI of all data and the data selected by threshold only. The selected data has the very high NMI, proving that our LGL can successfully filter the reliable pseudo labels. It is noted that we perform this analysis after we complete all the experiments and do not use it to guide the self-supervised training process.


\newpage
\balance
\footnotesize
\bibliographystyle{IEEEbib}
\bibliography{refs}
\end{document}